\documentclass{svmult_aa}

\usepackage{mathptmx}       
\usepackage{helvet}         
\usepackage{courier}        
\usepackage{type1cm}        
\usepackage{graphicx}        
\usepackage{multicol}        
\usepackage[bottom]{footmisc}

\usepackage{natbib}

\begin{document}

\title*{Embedded Clusters}
\author{Joana Ascenso}
\institute{Joana Ascenso \at CENTRA, Instituto Superior T\'ecnico,
  Universidade de Lisboa, Av. Rovisco Pais 1, 1049-001 Lisbon,
  Portugal {\it and} Universidade do Porto, Departamento de Engenharia
  F\'isica da Faculdade de Engenharia, Rua Dr. Roberto Frias, s/n,
  P-4200-465, Porto, Portugal, \email{jascenso@fe.up.pt}}
\maketitle

\abstract{The past decade has seen an increase of star formation
  studies made at the molecular cloud scale, motivated mostly by the
  deployment of a wealth of sensitive infrared telescopes and
  instruments. Embedded clusters, long recognised as the basic units
  of coherent star formation in molecular clouds, are now seen to
  inhabit preferentially cluster complexes tens of parsecs
  across. This chapter gives an overview of some important properties
  of the embedded clusters in these complexes and of the complexes
  themselves, along with the implications of viewing star formation as
  a molecular-cloud scale process rather than an isolated process at
  the scale of clusters.}



\section{Introduction}
\label{sec:introduction}

The study of embedded clusters dates back to the first infrared
detectors for astronomical use. Still enshrouded in the dusty
environment of their natal molecular cloud, embedded clusters are
invisible to optical telescopes but reveal themselves as rich and
fascinating objects at longer wavelengths. They contain the youngest
stars formed, and are therefore invaluable probes of the star
formation process. Their stars share the initial conditions of their
parent clump of gas, inheriting some of its characteristics, later
probed by humans in an attempt to understand the sequence of events
dominated by the interplay between gravity, turbulence, and magnetic
fields that ultimately forms them.

Both observations and theoretical simulations of star formation have
grown in number and in detail since the seminal review of
\citet{Lada03} on embedded clusters. Observationally, the largest
leaps forward were the widespread shift from the study of individual
embedded clusters to the larger context of their molecular clouds, and
the large sky surveys to build an increasingly complete census of the
star formation in the Galaxy. Also important, the detailed study of
extreme star formation events, even by Milky Way's standards, has
expanded the parameter space for studies of star formation to the
limit of extragalactic studies.  These advances were made possible at
such a large scale by the deployment of near- and mid-infrared
telescopes and instruments, both in ground-based and in space
observatories. The Two Micron All Sky Survey \citep[2MASS,
][]{Skrutskie06}, that covers the entire sky, and later the {\it
  Spitzer Space Telescope} were invaluable at revealing the detailed
intricacies of entire star forming regions as well as to allow a
multitude of large scale surveys. {\it Spitzer} legacy programs such
as the Cores to Disks \citep[c2d, ][]{Evans03}), the Galactic Legacy
Infrared Mid-Plane Survey Extraordinaire \citep[GLIMPSE,
][]{Churchwell09,Benjamin03}, and the MIPSGAL \citep{Carey09}
programs, as well as dedicated surveys of individual regions, have
greatly advanced our understanding of star forming regions, producing
numerous catalogues, most of which yet to be fully explored.
Ground-based observatories have also contributed significantly with
near-infrared telescopes used for surveys (e.g., 2MASS, UKIRT, ESO
VISTA), and with near-infrared adaptive optics assisted instruments
for deep and high-resolution studies of individual regions (e.g.,
GEMINI, VLT). In the far-infrared, the {\it Herschel Space
  Observatory} \citep{Andre05} is currently providing invaluable
insight into the youngest stages of star formation, bridging the gap
between the study of pre- and proto-stellar molecular clouds with
sub-millimeter and radio telescopes, and the study of embedded
clusters at NIR wavelengths. On the opposite end of the spectrum,
sensitive X-ray observations of star forming regions, made possible
greatly through the {\it Chandra X-ray Observatory}, have strongly
contributed to the effort of assessing the stellar populations of star
forming regions.

This chapter provides an overview of the observable properties of
embedded clusters in the important context of their molecular clouds,
brought to light by this massive technological development. The
analysis is limited to Galactic regions - those that can be studied in
greater detail -, and does not include the interesting star formation
taking place at and around the Galactic Centre; the reader is referred
to the review by \citet{Longmore14} for the latter. Section
\ref{sec:definition-emb-cluster} of this chapter elaborates on the
difficulty of adopting one single definition of ``cluster'' for all
studies of star formation, reviewing the most common definitions in
the literature, and what they entail. Section
\ref{sec:morphology-structure} reviews the observed structure and
morphology of embedded clusters and star forming regions, highlighting
the trends that have emerged from the increasing sample of studied
clouds, and what they reveal in terms of the underlying processes at
play. Section \ref{sec:SF-history} describes the constraints on the
timescales for star formation, crucial in any theory of star
formation, derived from the observations of the ages and age
distributions in embedded clusters and cluster complexes.

Other very interesting topics could be addressed in detail in the
context of embedded clusters, and are only mentioned briefly in this
chapter. The stellar mass distributions in clusters and on the
molecular cloud scale can reveal important properties of the star
formation process; the universality of the initial mass function, and
whether or not embedded clusters are mass segregated have been the
subject of many interesting studies in the past decade; the
consequences of the clustered environment to individual forming stars
at different stages of their evolution, and in particular their
formation along with massive stars is also an active topic of
research, and one that can help understand the probability of a given
star developing planets with certain characteristics. The analysis of
the efficiency and of the rate of star formation, both at the embedded
cluster and at the molecular cloud scales, is also starting to be
possible at great detail for a statistically significant sample of
known regions in the Galaxy. The topics included in this chapter are a
naturally biased selection of the what the author considers the most
robust observational advances in the last decade and most susceptible
of providing solid constraints to existing theories.

\section{What is an Embedded
  Cluster?} \label{sec:definition-emb-cluster}

An embedded cluster is a group of young stars that is still embedded
in its natal molecular cloud. Although seemingly simple, this
definition is all but trivial. The definitions we adopt reflect and,
at the same time, somehow limit our understanding of star
formation. Let's start with the definition of ``embedded'' and then
move on to the definition of ``cluster''.

\begin{figure}
\sidecaption
\includegraphics[width=8cm]{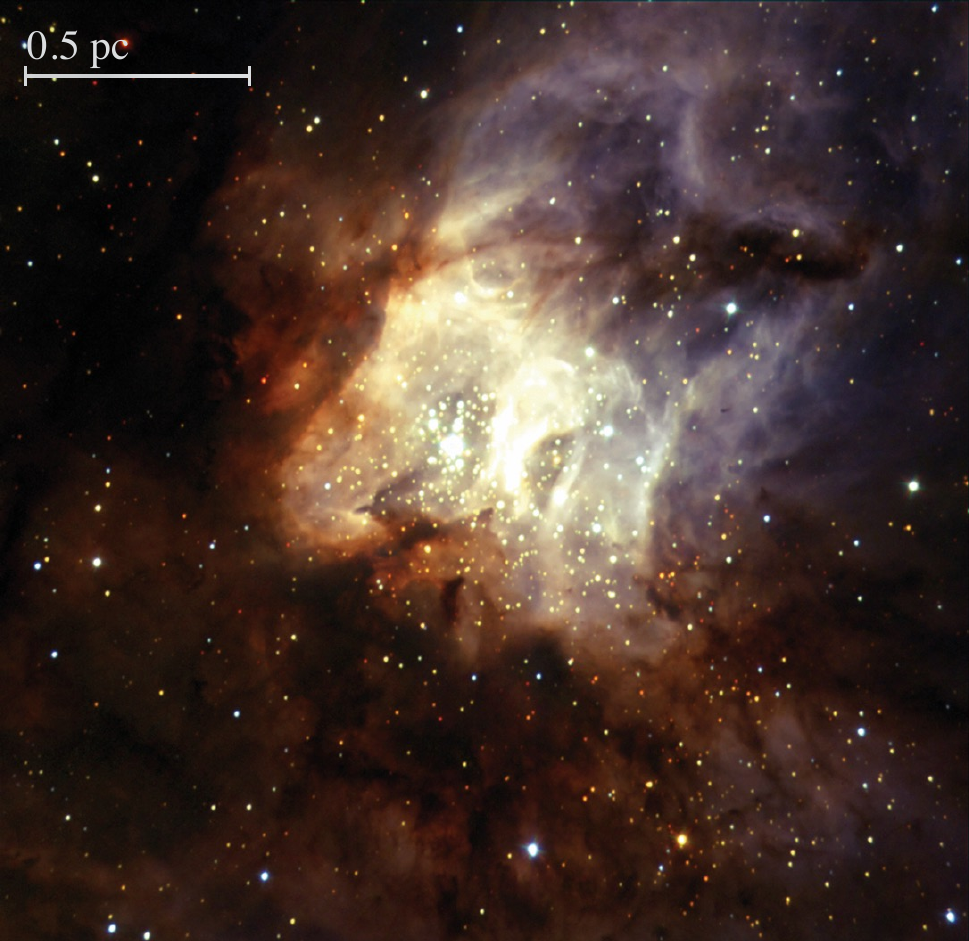}
\caption{RCW 38 is a young embedded cluster, imaged here in the
  near-infrared bands $J$, $H$, and $K_S$ with ESO/NTT/SOFI (Ascenso,
  Alves et al).}
\label{fig:rcw38}
\end{figure}

\subsection{Defining ``embedded''}
\label{sec:def-embedded}

An embedded star (or cluster) is one that is still enshrouded in its
natal molecular cloud. It is typically not (fully) observable at
optical wavelengths due to the heavy obscuration caused by the dust
grains in the cloud, but it can be seen in the near-infrared, where
young stars emit significantly \citep[e.g.,][]{Adams87,Robitaille06},
and the dust is more transparent
\citep{Savage79,Cardelli89,Rieke85,Draine11}. Near-infrared telescopes
and instruments are therefore the choice of excellence to detect and
characterise embedded objects, and indeed both ground-based and space
telescopes equipped with infrared detectors and filters have boosted
our demographics and our understanding of embedded clusters
exponentially in the past three decades.

It should be noted, perhaps trivially, that not all heavily obscured
objects are embedded: there are objects that are just seen behind
molecular clouds, and are therefore not within them \citep[e.g.,
][]{Alves01}. Objects that are in fact embedded notoriously display
signatures of youth. Since stars tend to disperse their natal gas and
dust via accretion and feedback over time an embedded star or cluster
is one that is necessarily young, and this leads to some unspoken
confusion regarding the ``embedded'' nature of clusters.

The canonical timescale for a cluster to clear enough material to
become optically visible is around 5 Myr \citep{Leisawitz89}, although
more recently \citet{Morales13} analysed the association of several
young clusters with molecular material, and proposed an upper limit of
the embedded phase of 3 Myr, while \citet{Portegies-Zwart10} quote a
duration of 1 to 2 Myr for the embedded phase of a cluster. But a
cluster's embedded phase should be a sensitive function of the mass of
the stars being formed. For example, massive stars develop HII regions
that are much more efficient in dispersing the cloud material than the
outflows from low-mass stars \citep{Matzner02}, so clusters with
massive stars should be the fastest to clear their surroundings and to
emerge from their molecular clouds. Therefore, although the condition
of being embedded is enough to attest to an object's youth, it is, by
itself, a poor criterion for a sample of clusters of uniform age.

On some accounts, the definition of ``embedded'' is narrowed to refer
to a state when the potential of the cluster is dominated by the mass
of the molecular cloud \citep{Gutermuth09}, according to which many
known young clusters can no longer be considered embedded. Trumpler
14, Westerlund 2, and NGC3603, for example, are all believed to be
well under 5 Myr old, but even though they are still partially
obscured by cloud material, they have already cleared most of their
intracluster gas. So these clusters are embedded only in the sense
that they are still associated with the molecular cloud, since their
gravitational potential is no longer dominated by the gas.

For the purpose of this chapter we will focus on clusters that are
younger than 5 Myr and still associated with their molecular clouds,
regardless of their potential being dominated by the gas.

\subsection{Defining ``cluster''}
\label{sec:def-cluster}

The definition of ``cluster'' is more controversial, and it is
non-trivial for many reasons. The need to define ``cluster'' arises in
several different contexts, each focused on different aspects. In the
context of large-scale observational surveys, for example, a set of
uniform criteria is paramount to detect (new) clusters against the
field of the Galaxy in an automated yet robust way. When analysing the
birth conditions and the evolution of clusters over time, the most
useful criterion is probably their dynamical state. Depending on the
question one is trying to address, the physical aspects that are
considered relevant - and that should therefore be used to define
clusters as entities - may vary. Additionally, the details that
numerical simulations of star forming molecular clouds are
increasingly capable of producing raise the pressure to find
observable signatures of some key property of young stellar
populations that can be tied to a dominant physical process. It is
therefore not surprising to see several definitions of ``cluster'' in
the literature, nor that they evolve alongside with the progress in
our numerical capabilities.

Previous to any definition of cluster, one practical difficulty arises
already in finding the stars that actually make up a population, since
knowing whether a given star is physically associated with its
neighbours or if it is only co-located in projection is challenging,
especially for more evolved populations like open clusters. Stars
younger than a few million years offer the advantage that they share
properties that are distinguishable from older stars, providing
important clues to their membership \citep[e.g.,
][]{Lada87,Shu87a,Adams87,Gutermuth09,Meyer97,Feigelson99,Feigelson10}.
Observational studies of clusters therefore often start by identifying
the young stellar objects (YSOs), usually by analysing their
near-infrared colours and/or X-ray properties, and then proceed to
finding over-densities that qualify as clusters by some measure. The
cloud material associated with embedded clusters in particular
effectively blocks a fraction of background stars, partially filtering
out stars unrelated to the cluster and increasing the local stellar
density contrast.

Low density groups are sometimes distinguished from clusters and
classified as O(B) associations if they contain O (and B) stars \citep
[e.g.,][]{Blaauw64}, T associations, if they only contain low-mass
stars \citep{Herbig62}, or R associations, intermediate between the
two and associated with bright, reflection nebul\ae\
\citep{van-den-Bergh66}. These classes overlap in many cases, and have
largely fallen into disuse over time. When no criteria other than an
overdensity of stars is used, the terminologies ``stellar aggregate'',
``stellar grouping'', or similar are also found. The concept of
``Correlated Star Formation Event'' was introduced by \citet{Kroupa13}
as an alternative to the concept of ``cluster''; it refers to all the
stars that were formed in one given star formation event over a
spatial scale of about one parsec, regardless of their spatial
distribution in a star forming region at present. These stars would be
coeval to within the duration of the star forming event. Although the
identification of such events observationally is limited by our
ability to determine individual stellar ages, this is an interesting
concept that is perhaps more meaningful in understanding the
progression of star formation in a cloud than the overdensity concept
of cluster.

\subsubsection{Morphological criteria}
\label{sec:morph-criteria}

Empirically, a cluster is an overdensity of physically co-located
stars. This definition is often used loosely to refer to all instances
of stellar groups. In this sense, detecting clusters can be as
straightforward as finding surface density peaks by eye on large-scale
images, with or without some additional criterion to minimise
contamination from spurious stellar density fluctuations. In the case
of young clusters, these criteria are usually a minimum number of
members, or the association with some tracer of youth, like outflows,
ionised gas, or molecular gas and dust, for example
\citep{Faustini09,Dutra00,Bica03a,Bica03b,Borissova11,Borissova14,Majaess13,Froebrich07}.

Quantitatively, several authors have defined several empirical
criteria, most often calibrated to detect previously known clusters in
blind surveys. \citet{Ivanov02}, for example, require a stellar
surface density contrast of at least 3-$\sigma$ above the galactic
background, and at least 50 members to claim the detection of a
cluster. Similarly, \citet{Kumar06} require a stellar surface density
contrast greater than 2-$\sigma$ above the local background, but a
minimum number of only 8 members. \citet{Carpenter00} requires that
the total number of stars within a closed 2-$\sigma$ surface density
contour exceeds a 5-$\sigma$ enhancement with respect to the expected
stellar background. \citet{Porras03} differentiate between
``clusters'' and ``groups'' based on whether a given region contains
more or less than 30 stars, respectively. Alternatively, in a
variation of the density-threshold algorithm, \citet{Gutermuth05},
following \citet{Casertano85}, use the distance to the Nth nearest
neighbour as a proxy for local density, eliminating the need to bin
the data spatially to produce density maps where to look for
enhancements.

\citet{Gutermuth09} devised a more sophisticated method to isolate
what they called ``cluster cores'' from co-spatial, extended young
stellar populations also associated with the cloud; they analyse the
separation between neighbouring stars using the minimum spanning tree
(MST) algorithm, and define the edge of a cluster core where the MST
branch lengths become larger than some critical
distance. \citet{Bastian07} employ the minimum spanning tree in a
slightly different way, truncating the separation between stars to a
maximum allowed distance to define clusters. \citet{Mercer05} detect
clusters using an algorithm that calculates the probability of a given
overdensity being an actual cluster and not a chance projection effect
considering the statistical distribution of the background field,
still based on geometrical and density enhancement arguments but also
on luminosity and colour criteria.

\citet{Schmeja11} compares the performance of a few different
algorithms in finding star clusters, and gives additional references
to works where the algorithms were applied. This author finds, as
expected, that strongly peaked clusters are easily detected by all
algorithms, whereas low contrast clusters can fall below the radar,
which reflects the ambiguity in the very definitions.

\subsubsection{Dynamical criteria}
\label{sec:dyn-criteria}

The previous definitions of clusters as overdensities of stars,
although powerful, lack physical grounds. A common physical criterion
to define ``cluster'' observationally is the inferred relative
stability of the stellar groups. \citet{Lada03} classify a group of
young stars as a ``cluster'' on the basis of its survivability against
tidal disruption up to the age of typical open clusters (100
Myr). According to this definition, a group of stars is considered a
cluster if it contains more than 35 members, and if its density is
higher than 1.0 M$_\odot$pc$^{−3}$; an {\it embedded} cluster is one
that is also ``fully or partially embedded in interstellar gas and
dust''.

In theoretical work and in numerical simulations of star formation,
``cluster'' is usually synonymous with {\it bound} group of
stars. This definition is useful because it simultaneously contains
important information about the molecular cloud from which the cluster
formed and about it's long-term survivability, and because it leaves
out any spurious overdensity of unrelated sources. It is also a {\it
  possible} definition in those contexts, since theory has all the
information about a given system under investigation, which is almost
never the case in the context of
observations. \citet{Portegies-Zwart10} \citep[see also][]{Gieles11}
distinguish between clusters (bound systems) and associations (unbound
systems) on the basis of their age with respect to the system's
dynamical time\footnote{The dynamical time is the time a typical star
  would take to cross the system
  ($t_\mathit{dyn}=R_\mathit{cl}/\sigma_V$). This is not to be
  confused with the system's relaxation time -- the timescale on which
  the system reaches equipartition of energy via two-body encounters
  --, which is much larger.}. A system whose age is, at present, a few
times its dynamical time has survived disruption by dynamical effects
for long enough to be considered a ``cluster'' according to this
definition. These systems are likely to survive as bound entities for
a significant fraction of a Hubble time \citep{Portegies-Zwart10}.

A dynamical analysis enables many interesting studies, including a
comparison between the molecular clouds and their stellar products:
systems (or subsystems) that are bound when they are very young are
likely to have formed monolithically from a bound, gravity dominated
cloud, whereas their unbound counterparts are more likely to have
formed from unbound, turbulence supported clouds. But the dynamical
state of a cluster is often difficult to assess, and one subject to
many uncertainties. In \citet{Portegies-Zwart10}, for example, the
definition of ``cluster'' depends strongly on the knowledge of the
cluster's age, of its mass, and of its virial radius. The
determination of a cluster's age from photometric surveys depends
mostly on the knowledge of the distance to the cluster, which can be
uncertain by a large amount for clusters that are too far away for
current measures of parallax; for example, the distance to the cluster
Westerlund 2 ranges from 2.8 to 8 kpc, even in the recent literature
\citep{Ascenso07,Carraro13,Zeidler15,Rauw11}. ESA's mission Gaia
\citep{Gaia-Collaboration16} will be an invaluable resource for
clusters that are already partially revealed in the optical. The
determination of a cluster's age (and age spread) is also importantly
sensitive to uncertainties in other properties like unresolved stellar
multiplicity, differential extinction between cluster members, and
stellar variability, including episodic accretion, and to the accuracy
of the stellar evolutionary tracks themselves
\citep{Hartmann01,Jeffries10,Preibisch12}. Estimates of cluster
masses, on their turn, can be severely affected by incompleteness,
poor membership assessment, or variable detectability over the
surveyed area due to, for example, extended, uneven bright nebula or
patchy extinction. Estimates of mass are also only as reliable as the
measurements of distance and age of the cluster, which, as outlined
above, are significantly uncertain. And depending on the wavelength,
they are more or less sensitive to the shape of the local extinction
law, and also to the specific pre-main-sequence evolutionary tracks
chosen to convert luminosity into mass. Finally, a cluster's virial
radius is taken as a factor of the half-light radius and assumes a
given stellar density profile. In rigour, only a spectroscopic
analysis of a significant fraction of cluster members at moderate
spectral resolution can determine their velocity distribution and
allow for a proper characterisation of a cluster's dynamical state,
but this is discouragingly expensive in observation time. As a
consequence, our knowledge of the dynamical state of the many known
clusters is still limited to an educated guess, and in particular it
is still too unreliable to be a strong observational constraint to
theories of star formation.

The very significance of the definitions of ``cluster'' based on
dynamical arguments inferred by observations has been called into
question by studies that suggest that there is no fundamental
difference between the stellar density distributions of ``clusters''
and ``non-clusters'' by any one definition. \citet{Bressert10}, for
example, do not find any bimodal signature in the stellar density
distribution of several star forming regions that suggests a preferred
or a threshold density for ``clusters'', although their sample
includes only a few clusters, of relatively low-mass, and their
diagnostics may be considered ambiguous \citep{Pfalzner12,Gieles12}.

In light of the previous arguments, it is clear that we are currently
not in position to make a statistically accurate comparison of bound
and unbound clusters, or of clusters and associations. At best, we can
attempt to rank known clusters in order of density, mass, luminosity,
or age, and try to find meaningful correlations that can be used to
constrain the physical conditions for star formation under different
environments.

In the context of this chapter, a ``cluster'' will be taken as its
most simple literary meaning: a collection of physically associated
stars.

\section{Morphology and Structure}
\label{sec:morphology-structure}

Embedded clusters come in a variety of forms. This can be inferred
instantly by comparing the images of a few star forming regions. It
was the striking morphological difference between different young
clusters that led to their traditional classification as``centrally
condensed'' or ``hierarchical'' \citep{Lada03}: the first refers to
clusters where the surface density has one strong peak and then
smoothly declines radially, and the latter to density distributions
with multiple peaks and a high level of sub-structure.

The importance of defining a cluster's morphology extends beyond the
need for uniform characterisation criteria. Rather, different
morphologies are produced by different conditions of the progenitor
cloud, they reflect different dominant physical phenomena, and they
can be predictive of the cluster's survival as bound entities on large
timescales or of their demise into field stars.

\subsection{Observational challenges}
\label{sec:observ-chall}

Similarly to detecting clusters, analysing their morphology has
important observational challenges. Incompleteness is the obvious
enemy of morphological studies: often only a relatively small fraction
of a cluster's members can be detected. The distance and the limited
sensitivity of instruments act against the detection of faint stars;
the limited resolution of the instruments acts against resolving
individual stars in a cluster, an effect that is additionally
amplified in very dense and/or distant clusters; the presence of
bright stars hampers the detection of less luminous neighbours out to
significant projected distances; and the interstellar extinction and
the bright nebula typical of star forming regions, which are almost
always variable in embedded clusters, change the detection limits and
the completeness spatially, producing artificial structure in the
observed distribution of cluster members. Also important is the
contamination from field stars, as mentioned before in section
\ref{sec:def-cluster}; unless cluster members are efficiently
distinguished from field stars, the analysis of their spatial
distribution can be significantly biased, especially in the case of
low surface density clusters.

Infrared observations can minimize some of these effects. Extinction
at longer wavelengths is significantly lower than in the optical
\citep{Rieke85}, providing deeper and more uniform completeness
levels. Also, the dynamic range of stellar brightness is lower in the
infrared than in the optical, i.e., the luminosity contrast between
the massive and low-mass stars will be smaller, making the latter
easier to detect.

\subsection{Cluster morphologies}
\label{sec:morphologies}

The human brain can readily distinguish between a centrally condensed
distribution and one that is more substructured, but an objective
measure of structure that can be applied uniformly to a large sample,
and one that can be quantitatively compared with results from
simulations and between different regions is required to build a
statistical framework for the properties of star forming regions.

Clusters visually recognised as centrally condensed are generally
relatively isolated clusters, with most members located in a
relatively small projected area in the sky. It is possible to define a
``centre'' for the cluster as the location of the maximum stellar
surface density, for example, and the surface density itself then
decays away from that centre as a smooth function in a way somewhat
resembling globular clusters. Analytically, the surface density decay
of a centrally condensed cluster is typically well described by a
simple power-law, by a power-law with a flat core \citep{Elson87}, or
by a King profile \citep{King62, King66}. The latter is parametrised
by the density at the cluster's core, by its core radius, and by a
tidal radius, and formally describes the density distribution expected
of a single-mass dynamically relaxed population that is tidally
truncated by an external (galactic) potential. While this is not an
accurate description of embedded clusters, the King profile is used as
a convenient function with few parameters, allowing for a uniform
description of the morphology of centrally condensed clusters
\citep[e.g.,
][]{Hillenbrand98,Ascenso07,Ascenso07b,Gutermuth08,Sung04,Wang08,Harfst10,Kuhn10}. The
\citet{Elson87} profile is often preferred in numerical simulations of
clusters, although it is also used to fit observed density profiles of
young clusters \citep{Brandner08,Gutermuth08,Gouliermis04,Sana10}.

Conversely, the stellar surface density of substructured clusters does
not follow a smoothly decaying radial function, instead showing
multiple peaks over some projected area. Several metrics have been
proposed to describe their fractal-like structure, including the
two-point correlation function \citep{Gomez93}, to describe the
probability distribution of any given star having a companion at
increasing distances, the distribution of mean surface density of
companions of cluster members \citep[][see also
\citet{Bate98}]{Larson95}, the normalised correlation length,
$\bar{s}$ \citep{Cartwright04}, defined as the mean
separation between cluster members normalised to the radius of the
cluster, and the normalised mean edge length, $\bar{m}$, of the
minimum spanning tree defined by the cluster
members. \citet{Cartwright04} review these methods in some detail
\citep[see also][]{Schmeja06}, and propose what they call the
$Q$-parameter as the most robust parameter to characterise the
morphology of a cluster. The $Q$-parameter is defined as the ratio
between $\bar{m}$ and $\bar{s}$, and is able to quantify the degree of
subclustering, as well as to distinguish between a centrally condensed
morphology and a hierarchical morphology: a $Q$ parameter larger or
smaller than 0.8 implies a large-scale radial density gradient or the
presence of subclustering, respectively. This parameter has since
become a widespread tool to analyse the structure of embedded
clusters.

It is worth noting that, in rigour, a substructured distribution of
stars, although commonly dubbed ``hierarchical'', is not necessarily
fractal. The loose classification of ``hierarchical'' in the context
of clusters usually refers simply to clusters with more than one peak
in stellar density, but \citet{Bate98} caution that the surface
stellar density distribution in a few known star forming regions
previously classified as fractal was also consistent with the stars
being distributed in random sub-clusters, a non-fractal
distribution. This distinction is important when interpreting
observations of cloud structure and stellar density distributions in
young clusters in light of the dominant physical processes, and also
when the number of stars is small enough that statistical fluctuations
can lead to the illusion of substructure.

\subsection{The molecular cloud scale}
\label{sec:mc-scale}

\begin{figure}
\sidecaption
\includegraphics[width=\textwidth]{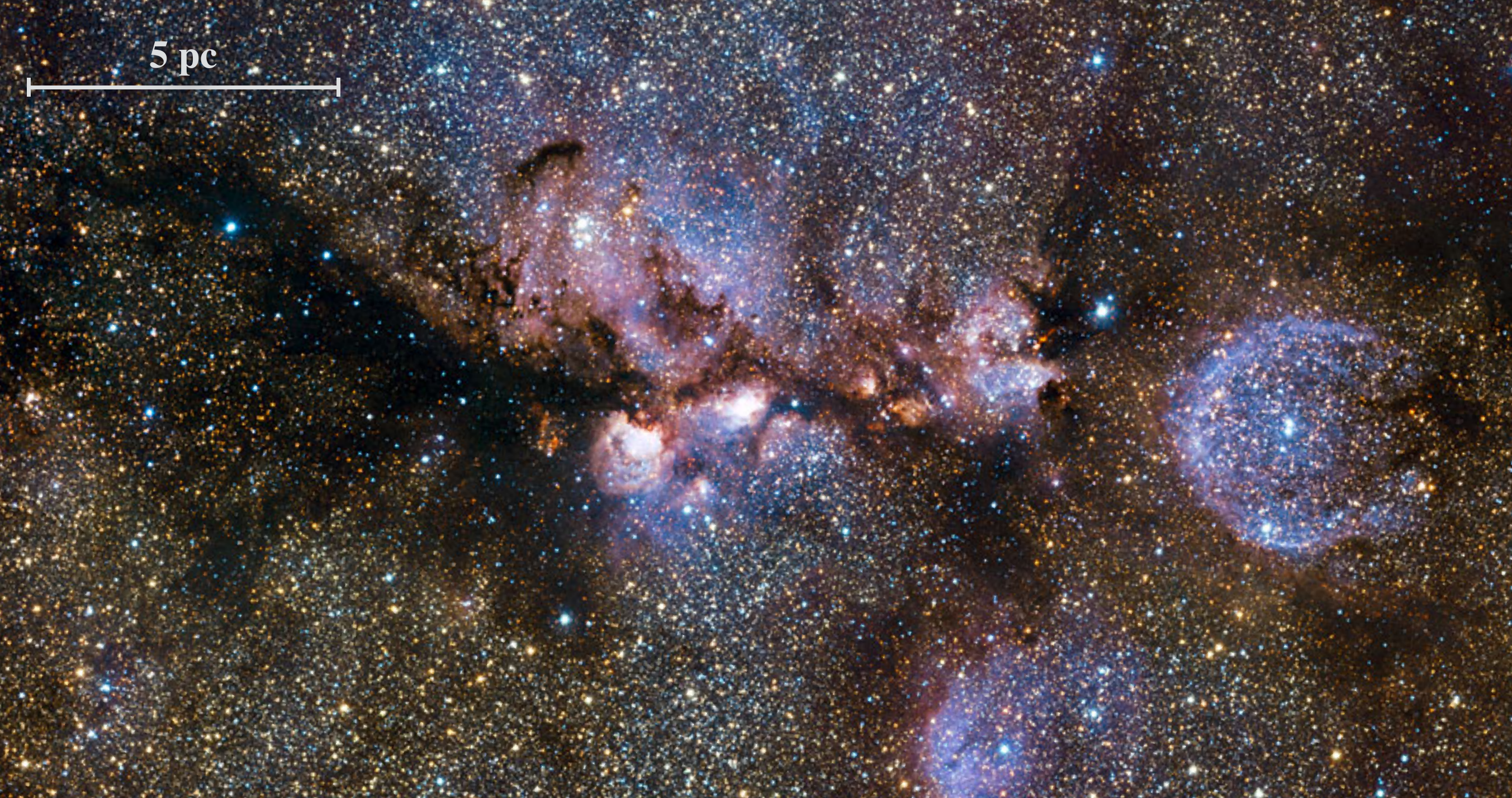}
\caption{The large-scale view of NGC 6334 imaged by ESO/VISTA in the
  near-infrared bands $J$, $H$, and $K_S$ (galactic North is up,
  galactic East is to the left, credit ESO/J. Emerson/VISTA). NGC 6334
  contains several embedded clusters along its actively star forming
  ridge.}
\label{fig:6334}
\end{figure}

The prolific effort to find new clusters in the Galaxy has already
yielded a sizeable database of embedded cluster candidates. Some
surveys target individual clusters, and are typically deep enough to
produce a comprehensive census of the stellar population down to the
low-mass end of the YSO mass spectrum; due to observational time
constraints and spatial resolution limitations, these surveys are
mostly limited to nearby, low-mass clouds, that harbour relatively
low-mass clusters as well. Other works encompassed observations of
entire molecular clouds, revealing interesting patterns of young
stellar populations. A few examples of deep surveys covering the
molecular cloud scale are the early works of \citet{Lada91,Lada92} and
\citet{Strom93}, for example, and the more recent dedicated surveys
of, e.g.,
\citet{Allen07,Carpenter00,Evans09,Roman-Zuniga08,Gutermuth09,Gutermuth11}
and \citet{Kuhn14}. On the massive end, only two surveys covered the
molecular cloud scale to a level comparable to more nearby star
forming regions: \citet{Preibisch14} and \citet[][see also
\citet{Wright14}]{Reipurth08} review the stellar population and the
clusters of the Carina and of the Cygnus X complexes, respectively,
each containing well over $10^4$ M$_\odot$ in young stars.

Blind, large scale or even full sky surveys provide more complete
censuses of embedded clusters at the galactic scale, necessarily
covering a wider range in cluster mass and different
environments. Even though so far most of the cluster candidates
identified in these surveys are not yet sufficiently characterised --
for most cases even the number of stars belonging to each cluster is
not yet properly assessed -- several tendencies have already began to
emerge, mostly supporting on a larger scale the understanding derived
from surveys of local star forming regions.

\subsubsection{Cluster complexes}
\label{sec:cluster-complexes}

\begin{figure}
\sidecaption
\includegraphics[width=\textwidth]{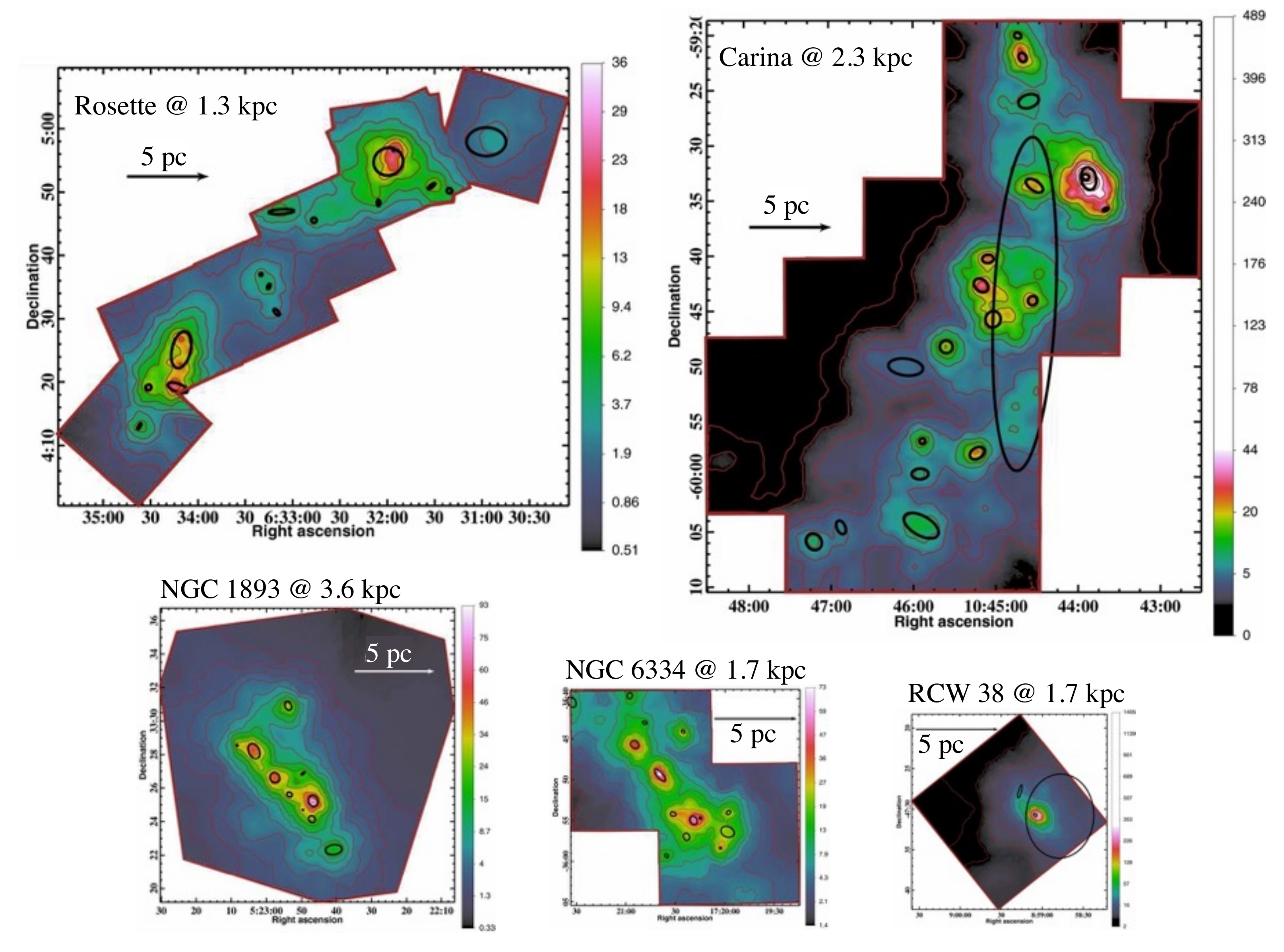}
\caption{Observed YSO surface density distributions for a few star
  forming regions registered to the same physical scale \citep[adapted
  from][]{Kuhn14}. These distributions illustrate well the cluster
  complex morphology in almost all regions that were observed in this
  work at the few-parsec scale. The colour bars are in units of stars
  pc$^{-2}$. }
\label{fig:ccomplexes}
\end{figure}

Surveys of individual molecular clouds have long suggested that star
forming regions are significantly substructured. Rather than
containing one single cluster with all or most YSO's, many nearby
regions contain several clusters organised in a more or less
hierarchical way (Fig. \ref{fig:ccomplexes}). A few well known
examples covering the low-mass end are Serpens and Perseus, Lupus, and
Chameleon (I and II); Orion, the Rosette Complex, Vela, the W3/W4/W5
complex, and RCW 106 are examples in the intermediate-mass range; and
among the most massive we know the Carina complex, Cygnus X, NGC 6334,
W51, W49A, that contain clusters that are more massive individually
than entire lower-mass cluster complexes (see several authors in
\citet{Reipurth08a,Reipurth08b}, and \citet{Evans09, Roman-Zuniga08,
  Nguyen15} for descriptions of these regions).

The same tendency is found in the most recent embedded cluster
catalogues that span wider ranges in heliocentric distance, and
presumably in mass; in the sample of \citet{Bica03a} 25\% of embedded
clusters have other clusters in their immediate (projected)
surroundings; \citet{Morales13} find that more than 50\% of the
clusters in their sample are in cluster complexes; \citet{Kuhn14} find
substructured distributions of YSOs in all of their targeted
clouds. In their sample of very young embedded clusters,
\citet{Kumar06} also find a strong tendency for complexes to show
substructure with 80\% of the clouds exhibiting multi-peaked surface
density distributions, already at very young ages; these authors
applied the same morphological classification to the relatively older
embedded clusters of \citet{Lada03} and found a similar
fraction. Although these numbers are not yet entirely reliable given
the incompleteness of these surveys, they suggest that the most common
outcome of star formation from molecular clouds is then cluster
complexes\footnote{I will refer to ``cluster complex'' as the global
  clustered YSO population within one cloud, and to ``cluster'' as the
  individual clusters within the complex.}, as opposed to single
clusters.

The size of these cluster complexes in the Galaxy varies from a few to
a few tens of parsecs along their largest dimension. The spread in
their clusters' size is smaller, around 1 pc
\citep[e.g.,][]{Kuhn14,Banerjee17}, mostly depending on the definition
of cluster size and on differing observational limitations. To some
degree, the distinction between a centrally concentrated and a
hierarchical stellar distribution can be regarded as a matter of
scale, as already hinted by \citet{Lada03}: at the tens of parsec
scale (cluster complex scale) substructure is ubiquitous, whereas at
the 1-pc scale (cluster scale) whatever observed substructure is
usually undistinguishable from statistical number fluctuations in a
centrally peaked, more or less elongated, distribution.

\begin{figure}
\sidecaption
\includegraphics[width=7.5cm]{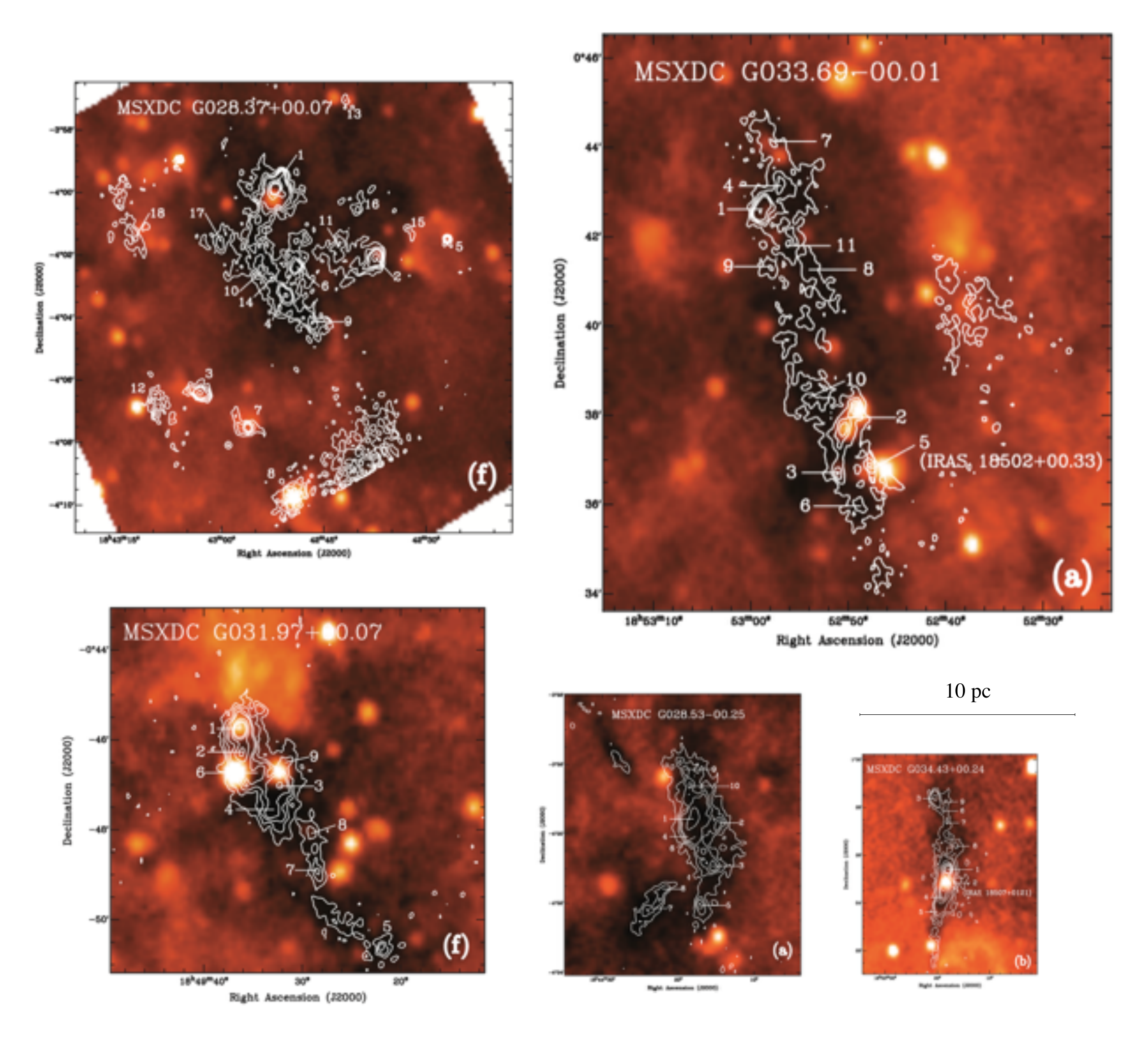}
\caption{Infrared dark clouds, presumably the precursors to clusters,
  often show elongated morphologies with large aspect ratios and
  multi-peaked density distributions over scales of $\sim$ 10 pc,
  similar to cluster complexes. Figure adapted from
  \citet{Rathborne06}.}
\label{fig:IRDCs}
\end{figure}

Overall, the distribution of young stars in cluster complexes is
reminiscent of the distribution of dense gas in molecular clouds
\citep[e.g.,][]{Lada96,Testi00,Gutermuth09}, both with respect to
their hierarchical structure and to their geometry. Like molecular
clouds \citep[e.g., ][see also
Fig. \ref{fig:IRDCs}]{Rathborne06,Peretto09,Churchwell09}, cluster
complexes have elongated morphologies with large aspect ratios. This
resemblance is expected if cluster complexes are younger than the
dynamical timescale for the clouds, otherwise they would have had time
to dissolve and take on more spherical geometries. At the cluster
scale, because it is smaller, there may have already been significant
dynamical mixing during the early embedded phase or even earlier, in
the gas phase \citep{Elmegreen06a}. Still, although the presence of
substructure in a stellar density distribution implies that the system
is not yet dynamically relaxed, some authors caution against taking
the similarity of cloud morphology and the distribution of YSOs at
face value, showing numerical simulations that produce hierarchical
distributions of YSOs that bear little resemblance to the original
distribution of dense gas \citep{Parker15}. Also, even though
substructure is typically interpreted as evidence of turbulence as an
important agent in driving the process of star formation,
\citet{Krumholz14} argue that a hierarchical distribution of YSOs does
not necessarily stem from turbulence-dominated initial conditions.

\subsubsection{Isolated clusters}
\label{sec:isolated-clusters}

Although the majority of star forming regions that have been studied
in detail exhibits a significant degree of substructure over scales of
the order of tens of parsecs (cluster complexes), there are a few
interesting exceptions -- single clusters that appear to be the sole
significant product of their natal molecular cloud. In the Galaxy,
excluding the peculiar vicinities of the Galactic Centre, a few
embedded clusters stand out as relatively isolated, as far as current
data suggests: Westerlund 2, NGC 3603, NGC 6611, and RCW 38 are a few
of those\footnote{A few other known clusters could be mentioned, such
  as W40, GM 24, or NGC 6618, for example, but the YSO populations of
  these clusters are not yet sufficiently well characterised to
  establish them as isolated in their clouds, or they are too close to
  other star forming regions that they may be part of a larger
  complex.}, and it is likely that more examples will emerge as the
new candidate catalogues start to be explored at higher detail with
state-of-the-art instrumentation. These clusters exhibit centrally
concentrated morphologies with faint hints of substructure at most,
and sizes less than, or of the order of 1 pc, similar to individual
clusters in the cluster complexes mentioned in the previous
section. However, their progenitor clouds do not seem to harbour other
clusters at present.

Low mass clusters are not considered in this context; since their
density contrast with respect to their surroundings is typically
small, any low-level extended population of young stars in the cloud
will provide comparable numbers of stars that they cannot be
considered isolated anymore. This introduces a bias that needs to be
kept in mind: the fact that the four isolated clusters considered here
are significantly more massive than the average individual cluster in
cluster complexes does not necessarily mean that isolated clusters
tend to be massive, nor that massive clusters tend to be isolated
(Carina is an excellent counter-example of the latter). We will come
back to these isolated clusters later.

\subsubsection{Unclustered young stars}
\label{sec:distr-trigg-SF}

\begin{figure}
\sidecaption
\includegraphics[width=7.5cm]{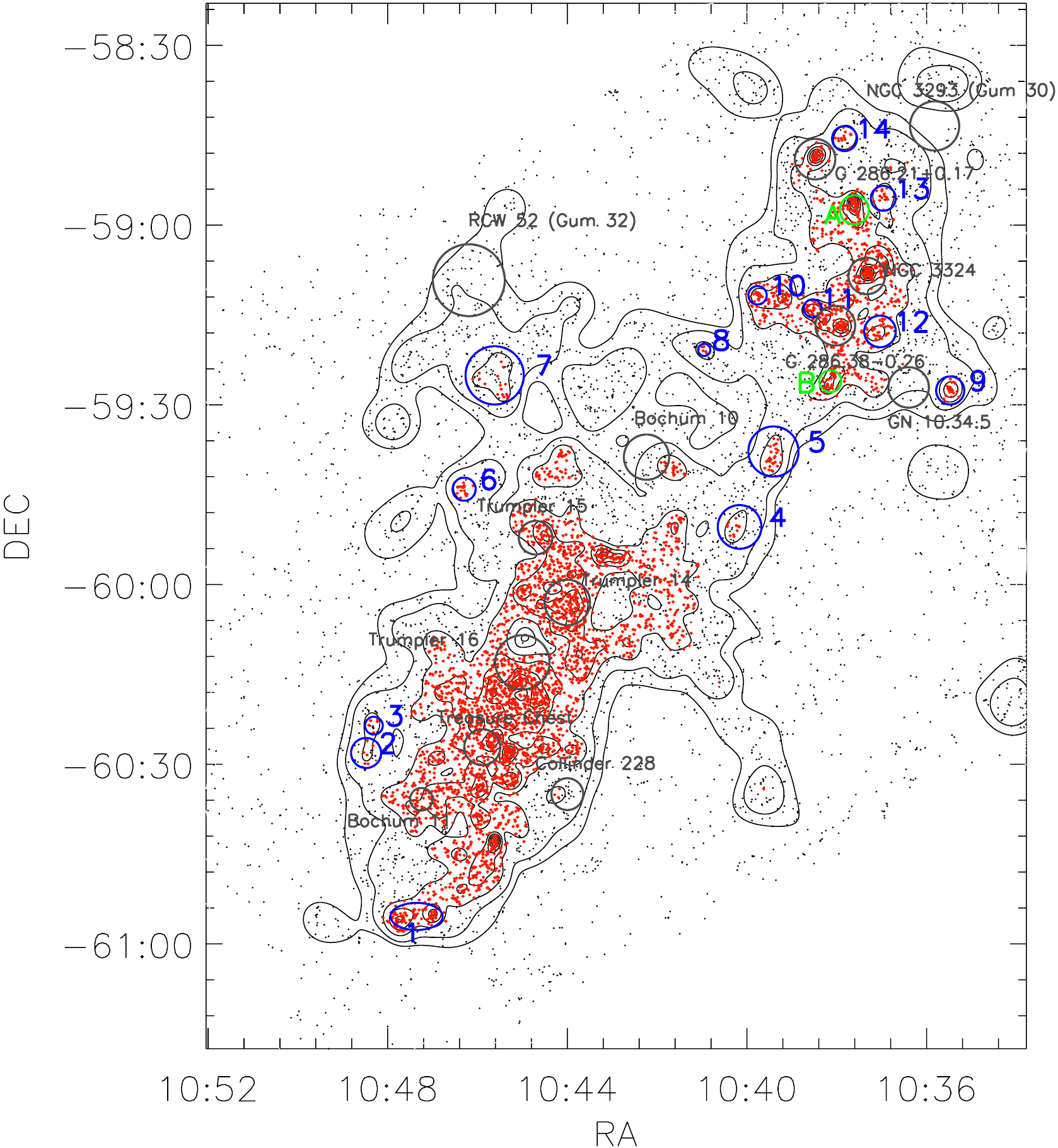}
\caption{YSOs are often found permeating entire star forming
  regions. This plot of the position ({\it left}) and surface density
  ({\it right}) of YSOs in the Carina Nebula from \citet{Zeidler16}
  shows a widespread population of unclustered YSOs throughout the
  complex.}
\label{fig:unclustered_carina}
\end{figure}

As implied above, not all young stars reside in the cores of embedded
clusters. Rather, a variable fraction of these stars is found
distributed throughout the embedding molecular cloud in relative
isolation (Fig. \ref{fig:unclustered_carina}). Large scale infrared
surveys, and later the {\it Spitzer Space Telescope} were instrumental
in showing that these distributed populations are ubiquitous in star
forming regions, most notably in cluster complexes. X-ray and infrared
combined YSO maps, less vulnerable to contamination from unrelated
sources albeit also less complete in particular mass ranges, confirm
the presence of widespread populations of young stars outside the main
clusters in star forming regions.

A reliable estimate of the actual fraction of isolated stars is
contingent on the definition of cluster and on several observational
parameters. To zeroth oder, accounting for a significant fraction of
the YSOs in a given region requires a sensitive sample with uniform
completeness limits, which is often challenging (see sect.
\ref{sec:observ-chall}). Also, since these objects are scattered over
large areas, observations should cover a large enough field of view
outside the main clusters, ideally covering the full extent of the
molecular cloud at comparable depth, which is observationally
expensive. It is equally important to accurately estimate the number
of stars that are {\it in} clusters, since underestimating this number
will enhance the weight of the extended population; this often
requires high resolution observations to adequately resolve the
crowded cores of dense clusters and account for the most of their
stellar population as possible. And finally, a reliable
de-contamination from field stars and distant galaxies is paramount,
since unrelated objects will artificially inflate the fraction of
distributed YSOs fairly easily. Once the young star population is
properly accounted for, the definitions of cluster and of the
boundaries of clusters obviously play an critical role in the
calculation of the fraction of stars that are outside clusters.

With this in mind, most estimates point to a relatively low fraction
of stars found outside clusters: in Orion A and B estimates are of a
maximum of 25\% distributed YSOs \citep{Allen07,Carpenter00}, around
the same fraction as for Ophiucus \citep[11-32\%,][]{Allen07} and
Perseus \citep[20\%,][]{Carpenter00,Jorgensen08,Evans09}; in Lupus and
in the Rosette complex, the fraction of YSOs found outside clusters is
estimated around 15\% \citep[][respectively]{Merin08,Roman-Zuniga08};
Monoceros R2 has a higher fraction of distributed YSOs, about 44\%
\citep{Carpenter00}. On the more massive end, in the W3/W4/W5 complex
more than 50\% of the stars are found in the five most massive
clusters; since the complex contains nineteen clusters in total, this
suggests that only a small fraction of YSOs is distributed
\citep{Carpenter00}; in the Carina complex an estimated 35\% of YSOs
is found outside the main cluster cores \citep{Feigelson11}, although
the number of cluster members could be underestimated in this
particular case since these observations cannot fully resolve the
highly crowded cores of the most massive clusters, significantly
underestimating the number of stars in these clusters. Surveys
including multiple star forming regions estimate an overall fraction
of ``isolated'' objects between 10 and 20\%, with upper limits of 40\%
\citep{Porras03,Koenig14,Gutermuth09,Evans09}.

The spatial distribution of these isolated stars in the cloud can be
useful in constraining their origin. They are often found to be spread
throughout the molecular clouds in a more or less uniform way, or, in
more quiescent clouds, still tracing the dense gas. These stars can
have formed at their current locations in relative isolation, they can
have been ejected from the nearby clusters, or they can be the
populations of slightly older clusters formed in the same cloud that
have already began to disperse away. A typically small fraction of
these stars is found in the nearby outskirts of clusters, toward
structures that were created by their feedback, for example at the
edges of bubbles or in pillars carved by the strong winds of the most
massive stars. Theoretically, stellar feedback is capable of
collecting and compressing existing molecular gas and create the
conditions for star formation in regions that would otherwise probably
not form stars, and this is likely the origin of some of the stars in
the distributed populations, but results from numerical simulations
suggest that this may account for only a small fraction. All these
scenarios produce stars with different ages compared to the stars in
clusters.

\section{Age Spreads}
\label{sec:SF-history}

As we have seen above, embedded clusters and star forming regions in
general are complex systems. It is not surprising that their histories
are also not simple. A molecular cloud does not form only one
generation of stars; rather, it is common to find populations
separated in age by a few million years associated with the same
molecular cloud, clearly suggesting that star formation does not occur
in a single burst and then stops. Understanding these age spreads,
which reflect the star formation history of the cloud, is fundamental
to understand the very process of star formation.

A review of the methods used to determine ages is beyond the scope of
this book, and the reader is referred to recent reviews \citep[][and
references therein]{Preibisch12,Soderblom10} for a discussion. It is
nevertheless important to mention that the determination of ages is
subject to many uncertainties, and that it is common for different
methods to return significantly different values. This is caused both
by observational limitations and by uncertainties in the
pre-main-sequence evolutionary models used to convert luminosities and
colours into ages and masses \citep[e.g.,
][]{Getman14,Jeffries10,Baraffe12,Preibisch12,Naylor09,Hartmann01,Burningham05,Hillenbrand08}. Using
synthetic clusters, \citet{Preibisch12}, for example, showed that a
coeval population of 3 Myr stars with the stellar variability, excess
emission from circumstellar material, and binarity fraction expected
for young stars, and subject to the differential interstellar
extinction typically found toward embedded clusters can present
near-infrared colours consistent with an age spread of more than 1
Myr.

For this reason the absolute ages inferred observationally for star
forming regions are still rather unreliable. Relative ages can be more
robust, as these are often inferred indirectly through the analysis of
the presence of circumstellar material. Circumstellar envelopes and
discs dissipate over time, such that the fraction of stars in a
cluster with circumstellar discs, for example, can provide a good
handle on the relative age of a cluster \citep{Haisch01,Briceno07}:
clusters with a large fraction of stars still with strong disc
emission are presumably younger than clusters where the majority of
stars is already discless. The characterisation of the emission from
the circumstellar material via spectral energy distribution (SED)
fitting \citep{Robitaille06} provides a finer age classification,
since the dispersal of discs follows a predictable logic. These have
the inconvenient that the timescale for the dissipation of discs is
mass-dependent, and that the fraction and characteristics of discs may
vary for the same age as a function of environment; for example, the
circumstellar material of stars that have close massive neighbours may
be affected by their strong feedback
\citep[e.g.,][]{Preibisch11,Johnstone98}. But in general SEDs allow
the distinction between younger and older pre-main sequence stars,
which, along with colour information and reasonably complete censuses
of the young stellar populations, is useful in constraining the
progression of star formation in a cloud.

Understanding age spreads in star forming regions is important at
several different scales, which again argues for surveys of entire
molecular clouds as important complements to narrower surveys of
individual clusters. On the scale of individual clusters, it is
interesting to assess the timescale over which their stars form,
whether individual clusters are formed rapidly, in a timescale
comparable to their dynamical time, or slowly and in quasi-equilibrium
\citep[e.g.,][]{Elmegreen00a,Tan06}; it is interesting to assess
whether they are formed monolithically already as large clusters from
a massive clump of gas, or are assembled from several
subclusters. These different scenarios require different conditions
from the progenitor cloud, and they operate under the influence of
different dominant physical processes, so they provide invaluable
constraints towards a predictive theory of cloud evolution and star
formation. At the scale of cluster complexes -- essentially the
molecular cloud scale -- it is interesting to understand whether a
cloud forms stars as a whole, or rather if different regions collapse
to form stars at different times; if the prompter for star formation
is internal or external to the cloud; if star formation develops
spontaneously from quiescent gas or if it is induced by some
event. Often neglected, the unclustered population distributed in the
cloud is intimately connected with the star formation on the clustered
scales, and its age distribution also contains important information.

The characterisation of age spreads and of star formation histories at
any scale is most meaningful in young regions for two reasons. First,
the relation of age with the fraction of stars with circumstellar
material becomes less sensitive for older populations, as stars
dissipate their discs. While the class 0/I phase is very short, around
0.5 million years \citep{Evans09}, class II and III stages last
longer, around a few million years. Second, given enough time,
dynamical processes will erase most of the imprint of the properties
of the progenitor molecular cloud on the stellar distributions,
decreasing the sensitivity of the analysis of ages and age spreads in
the context of their spatial distributions to the star formation
history of the cloud.

The term ``age spread'' will be used here to refer to the age
distribution of stars in a given context. We will review age spreads
within individual clusters, age spreads in molecular clouds, and age
spreads of the distributed/unclustered population of molecular
clouds. Some authors prefer the term ``age difference'' when referring
to the different ages of several clusters in the same molecular cloud,
reserving the term ``age spread'' to populations that have formed
together, in the same local event of star formation
\citep{Preibisch12}.

\subsection{Age spreads in cluster complexes}
\label{sec:age-spreads-clustercompl}

There is still not sufficient evidence to say whether different parts
of the same molecular cloud ``know'' about each other's status of star
formation. Depending on which phenomenon triggers star formation in a
cloud, it is possible that it occurs independently in regions that are
sufficiently far apart, that events of star formation are sequentially
triggered internally, or that the same trigger initiates star
formation in the cloud as a whole in a more or less synchronised
way. The differences are significant from the point of view of the
mechanisms at play, which means that studies of star formation benefit
greatly from analysing molecular clouds globally rather than only
individual regions within them. Cluster complexes in particular offer
a unique opportunity to study the progression of star formation in a
cloud, since each cluster can be viewed as a local event of star
formation within the common global history of the cloud.

\begin{figure}
\sidecaption
\includegraphics[width=\textwidth]{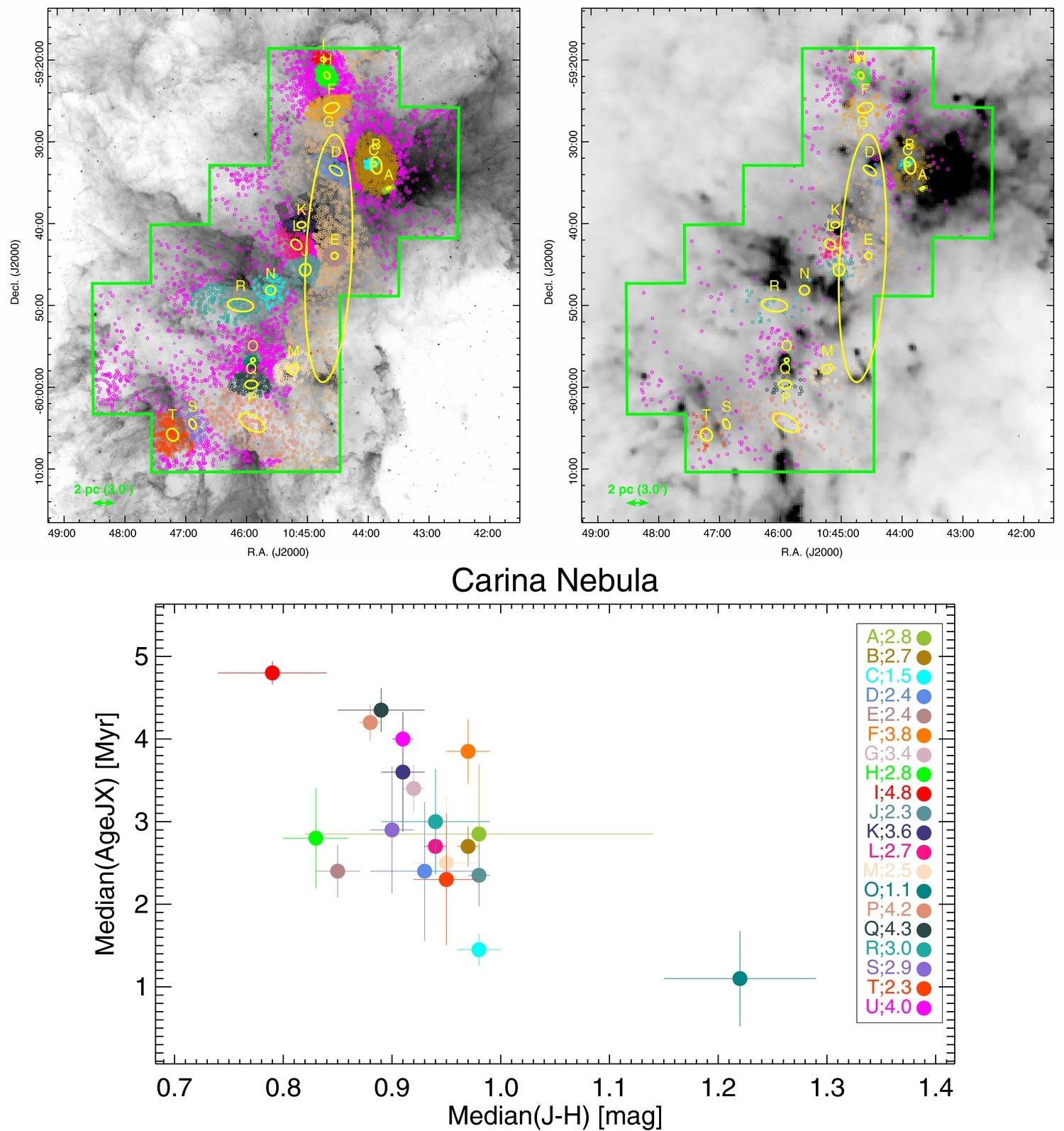}
\caption{Most cluster complexes harbour clusters with different ages
  but the global age spread is not too wide, nor is the age gap
  between any two clusters ordered chronologically. As in the Carina
  Nebula shown in this figure \citep[from][]{Getman14}, the spatial
  distribution of ages is often inconsistent with internal triggering
  being the dominant mechanism for the propagation of star formation
  within the cloud.}
\label{fig:age_spreads}
\end{figure}

Clusters of the same cluster complex often show different ages,
separated by as much as a few million years, as is illustrated in
Fig. \ref{fig:age_spreads} for the Carina Nebula. This is seen across
the mass spectrum of star forming regions. In the low-mass end,
\citet{Palla00}, for example, found age spreads larger than 3 Myr in
several nearby star forming regions (Taurus-Auriga, Lupus, Chamaeleon,
Upper Scorpius, and NGC 2264). Massive complexes, such as Orion,
Carina, or Cygnus, for example, show similar age
spreads. Interestingly, the maximum age gap between clusters of the
same complex if ordered chronologically is not too wide. There are a
few known examples of clouds that have ``very old'' (a few tens of
million years) and very young clusters with nothing in between
(Chamaeleon may be one such example), but most clouds show smaller
inter-cluster age gaps of the order of 1 Myr or less. In other words,
molecular clouds do not typically take long breaks between forming
clusters once they start, but they do not seem to collapse as a whole
either.

It is tempting to interpret the temporal proximity between different
clusters in the same cloud as evidence for sequential star formation,
with the first star formation event(s) triggering the formation of the
following, especially in clusters containing massive stars, those that
produce the most feedback. Although feedback can have a destructive
potential at small distances from the source star
\citep[e.g.,][]{Ngoumou15}, it can also collect and compress less
dense gas farther in the molecular cloud, or just precipitate the
collapse of pre-existing neighbouring clumps that would otherwise take
longer to, or never even, form stars and clusters
\citep{Elmegreen77,Bertoldi89,Whitworth94a,Dale07}. The exact
importance of these mechanisms as triggers for star formation depends
on the density distribution of the cloud prior to the influence of
feedback, and on the mass and location of the star(s) that produce the
feedback. The latter is particularly relevant because only stars
capable of producing an HII regions are able to trigger the collapse
of neighbouring clumps. In theory, the perturbation from the first
generation of stars is able to propagate and produce new stars (and
clusters) at the necessary speed across a typical cloud to reproduce
the observed age spreads at the observed distances between
clusters\footnote{For a 10 pc long cloud with a global age spread of 5
  million years, star formation would have to propagate at an average
  speed of at least 2 pc~Myr$^{-1}$, or 2 km~s$^{-1}$. This is about
  10 times the typical sound speed in molecular clouds assuming a
  temperature of 10 K ($c_S=(kT/m_{H_2})^{1/2}\sim0.2$
  km~s$^{-1}$). E.g., \citet{Getman14} suggest a propagation speed of
  star formation around 5 km~s$^{-1}$ in some of their clouds.}
\citep[e.g.,][]{Elmegreen77}, but in this scenario the spatial
distribution of ages at the scale of the cloud should show a coherent
progression. In the sample of star forming regions of \citet{Getman14}
-- the largest to date with stellar ages determined uniformly within
molecular clouds -- only about one third of the complexes show
reasonably coherent age gradients between clusters, suggesting that
internal triggering may not be the dominant controller of the
progression of cluster formation in molecular clouds.

Alternatively, an external event such as the passage of a spiral
density wave, nearby supernova events, or cloud-cloud collisions
\citep[e.g.,][]{Elmegreen98a,Cedres13,Dobbs09,Dobbs15,Dale15a,Fierlinger16}
could produce age distributions in clusters that are not necessarily
ordered, depending on the geometry and alignment of the cloud relative
to the triggering event. This would explain the lack of a coherent age
gradient mentioned above, and that some cluster complexes show no
significant age spread between clusters at all. For example,
\citet{Ybarra13} suggest that star formation started everywhere in the
Rosette Complex around the same time, suggesting that star formation
was somehow synchronised globally, presumably by an external
event. Also NGC 6334, the Cat's Paw Nebula, hosts a couple of slightly
older clusters, already partially revealed in the optical, and then a
molecular ridge spanning 10 pc of active star formation occurring in
discrete pockets at present \citep{Persi08}, challenging any
reasonable internally triggered star formation interpretation. The
external trigger scenario is attractive to explain such a large scale
coordination of star formation, although it is equally difficult to
prove. At some level it is not much different to discuss the formation
of stars at the molecular cloud scale and the formation of the density
structure in molecular clouds themselves, since it is not likely that
molecular clouds and dense clumps within them form spontaneously from
the interstellar medium. Considering that stars form everywhere there
is dense enough gas \citep{Lada10}, the problem of the progression of
star formation within a cloud is reduced to the problem of the
formation of molecular clouds.

At the molecular cloud scale, massive isolated clusters (see Section
\ref{sec:isolated-clusters}) are particularly interesting from their
age distribution perspective. These are apparently the sole products
of their molecular clouds: what is their history? Can they just be the
first generation of star formation in clouds that will later form
other clusters and host cluster complexes? Since the known clusters of
this type are already around 1 to 2 Myr old and their clouds do not
show evidence for substantial ongoing star formation, this scenario
would produce complexes with considerable age spreads, depending on
the timescale for star formation in different parts of the cloud. This
would not be unseen; the age spread in Chamaeleon between regions I
and III is likely larger than 10 Myr, and there does not seem to be
any cluster with an age intermediate between these two regions. But
this appears to be a rather atypical case. Carina, for example, has an
estimated age spread for clusters of $\sim$8 million years, between
Trumpler 15 and the Treasure Chest cluster \citep{Preibisch11,Smith05}
but small age gaps between consecutively formed clusters. Westerlund
2, classified above as one such isolated clusters, has a very young,
very embedded cluster forming just outside its borders, with hints of
massive star formation even, suggesting that clustered star formation
is still ongoing in its progenitor cloud. This is, however, the only
site of active cluster formation known in the cloud, which suggests
that this cloud is not likely to form a complex with many clusters, at
least not with a small age spread. Other clouds that contain clusters
as massive as Westerlund 2 (e.g., Carina, Cygnus, W49A) all contain
several similarly massive clusters, reinforcing the idea that
Westerlund 2 (and also NGC 3606 by similar arguments) is indeed
different from cluster complexes. RCW 38, the youngest of the three
isolated clusters considered here, also shows some evidence for
ongoing clustered star formation in its vicinities \citep{Winston11};
given its younger age (0.5 Myr) this cluster is more likely to evolve
into a (small) cluster complex than Westerlund 2 or NGC 3603.

The observation of cluster complexes is thus unveiling a non-obvious
scenario for the progression of star formation in molecular clouds. It
is clear that molecular clouds do not collapse globally but rather in
clumps that form clusters, and that there is no unique trend for the
age distribution of the clusters they produce. Also important,
molecular clouds seem to exhaust their star formation potential fairly
rapidly, on timescales of the order of a few Myr, the maximum age
spreads found in cluster complexes and the time in which they
typically disperse.

\subsection{Age spreads in individual clusters}
\label{sec:age-indiv-clusters}

Individual clusters refer here to clusters that have formed in a
single event, regardless of having formed alongside other clusters in
the same cloud. The Orion Nebula Cluster is an example of an
individual cluster in a molecular cloud (Orion A) that hosts other
clusters.

Most detailed studies suggest that the age spread in individual
embedded clusters is very small, typically within the age
determination uncertainties, if it exists at all
\citep[e.g.,][]{Preibisch02,Moitinho01,Jeffries11,Banerjee14,Banerjee15,Getman14}. The
main observational difficulty when analysing individual clusters --
more important even than the uncertainties from the age determination
method -- is contamination from stars that do not belong to the
cluster under study. Since clusters often reside in cluster complexes,
the contamination by stars from the complex may be significant, and
since the contaminant stars will also be young, distinguishing them
from a given cluster population can be difficult. As such, published
claims of significant age spreads within young clusters are often
challenged by subsequent larger scale surveys that reveal populations
of older YSOs spread out in the cloud, distributed more or less
uniformly far beyond the cluster's borders, suggesting that they are
not part of that cluster but are rather different populations within
the complex (see section \ref{sec:age-distributed-population}). If
these stars are taken as cluster members they will misleadingly
present as evidence for age spreads within the cluster.

The Orion Nebula Cluster is an example where considerable age spreads
have often been reported \citep{Palla99,Huff07,Da-Rio10}. The recent
study of \citet{Getman14} suggests a shallow radial gradient of
increasing age that is interpreted by the authors as the cluster
having formed outside-in. However, these results are equally
compatible with the cluster having a small age spread and being
immerse in a distributed population of older stars that do not belong
to the cluster: the cluster stars would bias the age toward younger
values in the centre, and the older, extended population would start
to weigh in toward the peripheries as it outnumbered the cluster
members. An older population, unrelated to the ONC but extending well
into its foreground, has indeed been found by \citet{Alves12}, which
could account for the older stars that make up the observed pseudo age
spread. On the more massive end, \citet{Ascenso07b} found hints of a
core-halo morphology in their small field survey of the cluster
Trumper 14, where the ``halo'' stars seemed older and appeared
unclustered; later, a significant population of older stars permeating
the entire Carina Complex was uncovered by a large-scale survey
\citep{Preibisch11}.

\begin{figure}
\sidecaption
\includegraphics[width=\textwidth]{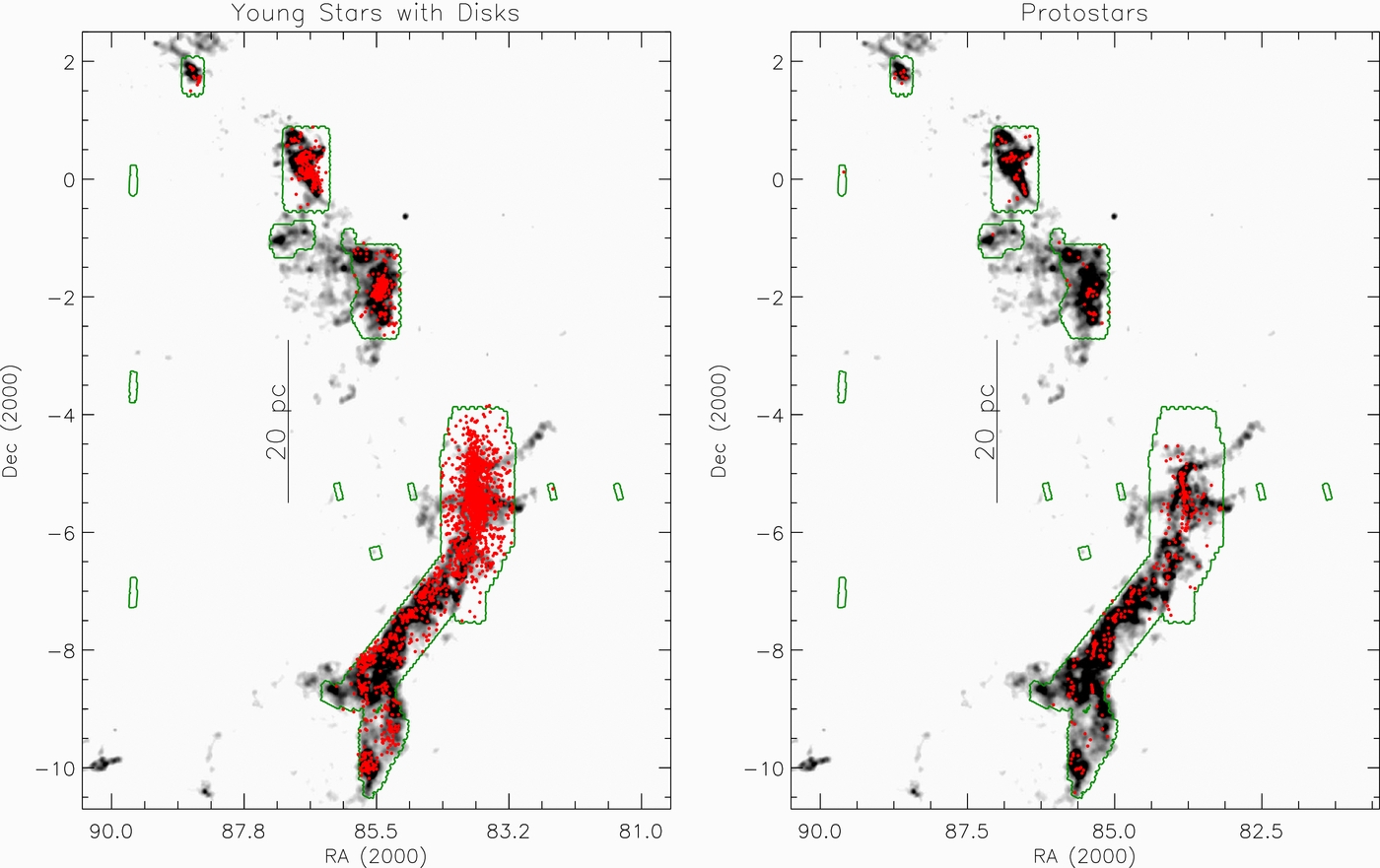}
\caption{Younger, class 0/I sources are typically found more tightly
  clustered than the older, class II sources. This is illustrated in
  this figure from \citet{Megeath12} of the Orion molecular clouds
  showing the distribution of protostars ({\it right}) and the
  distribution of older stars with discs ({\it left}).}
\label{fig:yso_class_dist}
\end{figure}
 
Indeed, studies of the population of YSOs in different stages of
evolution at the molecular cloud scale, made possible largely through
{\it Spitzer} observations, reveal this as a pattern: although young
sources are clustered in general (see section
\ref{sec:morphology-structure}), class 0/I YSOs -- those with the most
circumstellar material and the youngest -- consistently appear more
tightly clustered than their class II and class III counterparts.
Fig. \ref{fig:yso_class_dist} illustrates this typical distribution
for the Orion clouds. Since the class 0/I stage is very short-lived,
the position of these stars is very likely the position at which they
formed, supporting the view that stars form in dense configurations
within clouds \citep[e.g., ][]{Lada03}. The wider distribution of
class II and class III sources likely reflects the characteristics of
the unclustered population of stars in molecular clouds (see section
\ref{sec:age-distributed-population}).

Massive clusters are particularly interesting in terms of their age
spreads. They contain extraordinary numbers of stars, which means they
were formed from an extraordinarily massive gas clump or assembled by
mergers of smaller clusters. Observationally, some of the most massive
embedded clusters known in the Galaxy outside the Galactic Centre, do
not show evidence for significant age spreads
\citep{Kudryavtseva12,Stolte04,Ascenso07,Ascenso07b}. \citet{Banerjee14,
  Banerjee15} find that the observed properties of NGC 3603 in
particular are compatible with a starburst scenario and incompatible
with it having been formed through the coalescence of smaller
clusters. The suggested near-instantaneous, monolithic formation of
such massive clusters raises important questions regarding the support
of molecular clouds against gravitational collapse over long enough
timescales to assemble the necessary amount of dense gas (see section
\ref{sec:discussion}).

In light of existing evidence, individual clusters do not seem to have
significant intrinsic age spreads. Events that form individual
clusters seem to operate on very small timescales, of the order of, or
smaller than, the local dynamical times (less than 1 Myr for typical
clusters).

\subsection{Age spreads of the unclustered stars}
\label{sec:age-distributed-population}

The distributed stars found both in and around cluster complexes and
isolated clusters (see section \ref{sec:distr-trigg-SF}) show a wide
range in ages. Several of these populations\footnote{In this context,
  I use the term ``population'' loosely to refer to the collection of
  stars that are not clustered, without any implication regarding
  common properties or origin.} were found and studied in multiple
works using observations and methods sensitive to different ages, from
less than 1 Myr to 20 Myr. Adding to the results of the individual
studies, this diversity shows that the age spreads of these
distributed populations is rather large; and rather extreme as well:
the oldest, and often the youngest, stars in a star forming region are
found in the distributed population.

The youngest distributed stars are often found toward the edges of the
clouds, projected against shells, bright-rimmed clouds, or pillars,
that are illuminated and carved by the action of a cluster of slightly
older stars. This spatial correlation, sometimes backed by other
indicators, has been widely interpreted as evidence for star formation
triggered by existing stars or clusters. \citet{Dale15} compiled a
list of the many studies that have claimed observational evidence for
triggered star formation, most from positional arguments.

A small fraction of very young stars and protostars is sometimes found
along dense filaments of gas and dust in more quiescent regions of the
clouds, although not far from the location of the older stars and
clusters. These are not randomly distributed in the clouds, but apart
from their ordered location along the filaments, they are not
significantly clustered at the individual cluster ($\sim$ 1 pc) scale,
unlike the majority of stars of the same age.

Most of the distributed population of stars in a cloud is made up of
intermediate age pre-main-sequence stars (class II). Although they
usually also follow the overall clustering pattern of the star forming
region, they are typically less clustered than class 0/I objects. It
is not uncommon to find class II stars pervading the clouds at the
cluster complex scale (a few to a few tens of parsecs), both in
embedded and in less embedded regions. At this point it is useful to
note that class II (and III) sources can represent a wide range of
stellar ages, since more massive stars dissipate their circumstellar
material more rapidly, therefore acquiring the SED signatures
characteristic of class II YSOs at younger ages \citep{Williams11}. It
is therefore expected that (younger) class II sources be found
clustering with coeval class I sources, and that older class II
sources be found spread out in the cloud, as observed.

Wide distributions of old pre-main-sequence stars, as old as 20 Myr,
have also been reported in star forming regions dominated by younger
stars and clusters. In the Galaxy, Orion A, the Carina Complex, NGC
3603, and NGC 6611 in the Eagle Nebula all have reports of ``old''
populations in their clouds. Interestingly, in Orion A and in Carina,
these populations can tentatively be attributed to an identifiable
cluster, namely NGC 1980 and Trumpler 15, respectively, both
containing massive stars. Conversely, the ``old'' populations of NGC
6611 and of NGC 3603 have not been associated to any existing cluster,
although a giant molecular shell is observed in the Eagle Nebula that
can be the remnant of a supernova event \citep{Moriguchi02},
suggesting a possible association with the old pre main sequence
population.

\section{Stellar mass distributions}
\label{sec:imf}

This chapter would not be complete without a dedicated word about
stellar mass distributions in clusters. It is widely accepted that the
observed stellar mass distribution of a young cluster is a good
approximation of its initial mass function (IMF), and that this IMF
seems to be fairly universal across the spectrum of cluster properties
\citep[e.g.,][]{Lada03,Bastian10,Kroupa13}. On the theoretical side,
we have presently reached a stage where all accepted theories of star
formation are capable of producing the observed IMF of clusters,
undermining its predictive or constraining power. But recent and
upcoming observing facilities may change that, by changing the focus
of IMF studies slightly.

For example, it is not yet clear when exactly the IMF becomes fully
assembled, or whether massive or low mass stars preferentially form
first, or what impact, if any, the first formed stars have on the
formation of the subsequent population. In the future it will become
increasingly easy to study extremely young clusters, including of the
more distant massive clusters in the Galaxy, with adequate resolution
and sensitivity. Is the IMF of these clusters any different from that
of older clusters that have presumably already finished most of their
star formation activity, suggesting that different mass stars form at
different stages?

Also, it is only apparently clear that the IMF is indeed universal in
all environments. The same IMF is found in most star forming regions,
but some ``regions'', especially the less massive, include stars from
large physical volumes, sometimes from entire clouds, whereas others
refer only to individual clusters at the 1-pc scale. It is not clear
how these similar IMFs over such different scales can be made
consistent. As more and more cluster complexes are studied it will
become increasingly possible to assess the mass distribution of the
entire stellar population formed by one cloud with respect to the IMF
of the individual clusters, and to the IMF of the distributed
population. We must then understand what is the meaning of an IMF at
the molecular cloud scale. If different star formation events
(clusters) in the same cloud (cluster complex) are independent from
each other, then so should their IMFs, otherwise star formation must
be set at the global scale of the cloud rather than locally, reducing
the distance between studies of stars in clusters and cluster
complexes and studies of molecular clouds and assembly of dense gas,
towards a consistent picture of star formation.

\section{Embedded Clusters and Star Formation}
\label{sec:discussion}

Clusters and cluster complexes reveal intricate and often puzzling
star formation histories in molecular clouds. Observational results
suggest that star formation is a rapid and likely discontinuous
process at the molecular cloud scale. Rather than forming one cluster,
each cloud typically forms multiple clusters over timescales of a few
million years. Individual clusters themselves appear to be mostly
coeval, but around and between them significant populations of stars
with wide age spreads are often found. What can these spatial and age
distributions tell us about the origin and the progression of star
formation in clouds?

Different possibilities considered by theory and reproduced by several
flavours of numerical simulations predict different properties for
star forming regions that are becoming increasingly possible to
compare with observations. To this end one important step has been
taken in the last decade: more and more star forming regions are being
studied at the molecular cloud scale. The structure and age
distributions of young stars in molecular clouds are particularly
relevant in constraining the timescales for star formation, both
locally and globally, indirectly favouring one or other aspect of the
theoretical possibilities.

Individual embedded clusters span a wide range in mass and density,
but there is very little convincing evidence that they have large age
spreads (see section \ref{sec:age-indiv-clusters}). Individual
embedded clusters younger than $\sim$1 Myr are common, which suggests
that clusters are formed fairly rapidly, on timescales comparable to
their dynamical times. Their smoothly peaked morphologies at the
$\sim$1 pc scale already at these very young ages suggest that they
were formed from a molecular cloud clump that was itself already dense
with a peaked density distribution, or that any initial substructure
must have been erased very efficiently. The latter would argue for a
slower process of star formation that would allow time for dynamics to
act on pre-existing structure, but large scale observations of pre- or
proto-stellar clumps in massive, infrared dark clouds, presumably the
precursors to embedded clusters, often show individual clumps about
the size of embedded clusters already with fairly symmetric density
distributions \citep{Shirley03,Ragan12,Traficante15}. A large fraction
of these clumps shows signs of star formation, supporting further the
view that the starless phase of a dense molecular clump is very
short. Taken at face value, this and the small age spreads in
individual embedded clusters require that, for each cluster, a
significant amount of dense gas be gathered prior to the onset of star
formation, and that it does not fragment significantly in the
process. This may require a support against gravitational collapse
until conditions are met that precipitate the quasi-instantaneous
formation of a whole cluster of stars, especially for the most
massive; or, alternatively, this could be achieved if the dense gas
itself was gathered by a rapid phenomenon, such as collisions between
molecular clouds, collisions of filaments within molecular clouds, or
through the action of external agents, such as supernovae.

Cloud-cloud collisions have been recently invoked to explain the rapid
formation of massive clusters such as NGC 3603 and Westerlund 2. Based
on radio kinematic data, \citet{Furukawa09} and \citet{Fukui14} find
that each of these clusters lies at the interface between two massive
molecular clouds that seem to be moving towards each other with
relative velocities of $\sim$ 20 km/s. Hydrodynamical simulations
confirm that cloud-cloud collisions can form bound and massive clumps
and cores \citep{Habe92,Anathpindika10,Inoue13,Wu15}, but studies of
the characteristics of the produced stellar population are still
necessary to show that this mechanism is capable of forming entire
(massive) clusters. The same type of kinematical signature is found in
clouds harbouring lower mass and more substructured star forming
regions, such as M20 \citep{Torii11} and RCW120 \citep{Torii15},
suggesting that this mechanism, if indeed capable of forming clusters,
can reproduce a range of observed properties. This scenario is
particularly appealing in the cases of isolated massive young
clusters, where the gathering of the required amounts of dense gas is
particularly challenging.

Competing theories, complete with numerical simulations, posit that
clusters may be assembled hierarchically, with stars forming along
filaments and then falling to the deepest part of the potential well,
forming a cluster
\citep{Bonnell03,McMillan07,Bate09,Maschberger10}. Filaments are a
distinct characteristic in all molecular clouds, and young stars
within them are also ubiquitous in star forming regions, especially in
low-mass environments, lending support to this scenario. These
simulations do not require a mechanism to assemble massive clumps of
gas prior to star formation, and they also form clusters very rapidly,
although the actual duration of the star formation event depends
sensitively on the initial conditions. As a by-product, very extended
haloes of stars must form from stars that are ejected from the cluster
core through dynamical interactions as the subclusters merge
together. This could provide a natural origin for the extended
population of young stars that is very often found in star forming
regions (sections \ref{sec:distr-trigg-SF} and
\ref{sec:age-distributed-population}), and an overall consistent
picture for the formation of all stars in star forming
regions. However, the age of the extended population should be
consistent with the (narrow) age range of the final clusters, whereas
the majority of the distributed stars is often older than the
clustered population. Unless, since ages are often inferred through
the presence of circumstellar material, the ejection process strips or
truncates the discs from these stars, making them appear older to such
age diagnostics. For lack of computational power, it is also not yet
clear that numerical simulations that form clusters via hierarchical
assembly can produce clusters as massive as the most massive observed,
or that they can reproduce the larger scale cluster-complex morphology
prevalent in clouds with the observed age spreads under realistic
initial conditions.

\vspace{0.5 cm}

\acknowledgement Many thanks to Jo\~ao Alves for his encouragement and
for discussions that contributed to this manuscript. This work was
supported by FCT -- Funda\c{c}\~ao para a Ci\^encia e Tecnologia,
Portugal (grant SFRH/BPD/101562/2014 and FCT contract
UID/FIS/00099/2013).

\bibliographystyle{hapj}
\bibliography{/Users/jascenso/Research/Documents/new_bib}
\end{document}